\documentclass[twocolumn,prl,aps,showpacs,superscriptaddress]{revtex4-1}
\usepackage{graphicx,graphics}
\usepackage{amsmath}
\usepackage{amssymb}
\usepackage{epstopdf}
\usepackage{amstext}
\usepackage{amsthm}
\usepackage{amsfonts}
\usepackage{latexsym}
\usepackage{array}
\usepackage{xfrac}
\usepackage{color}
\usepackage[toc,page]{appendix}
\usepackage{subfigure}
\newcommand{\igbjd}[1]{}\newcommand{\beqa}{\begin{eqnarray}}
\newcommand{\eeqa}{\end{eqnarray}}
\newcommand{\beq}{\begin{equation}}
\newcommand{\eeq}{\end{equation}}
\usepackage[colorlinks,bookmarks=false,citecolor=blue,linkcolor=red,urlcolor=blue]{hyperref}

\begin{document}

\title{Entanglement entropy in low-energy field theories at finite chemical potential} 

\author{Ivan Morera}
\affiliation{Departament de F\'isica Qu\`antica i Astrof\'isica, 
Facultat de F\'{\i}sica, Universitat de Barcelona, E--08028
Barcelona, Spain}
\affiliation{Institut de Ci\`encies del Cosmos, Universitat de 
Barcelona, ICCUB, Mart\'i i Franqu\`es 1, Barcelona 08028, Spain}
\author{Ir\'{e}n\'{e}e Fr\'{e}rot}
\affiliation{ICFO-Institut de Ciencies Fotoniques, The Barcelona Institute 
of Science and Technology, 08860 Castelldefels (Barcelona), Spain}
\affiliation{Max-Planck-Institut f{\"u}r Quantenoptik, D-85748 Garching, Germany}
\author{Artur Polls}
\affiliation{Departament de F\'isica Qu\`antica i Astrof\'isica, Facultat 
de F\'{\i}sica, Universitat de Barcelona, E--08028 Barcelona, Spain}
\affiliation{Institut de Ci\`encies del Cosmos, Universitat de 
Barcelona, ICCUB, Mart\'i i Franqu\`es 1, Barcelona 08028, Spain}
\author{Bruno Juli\'{a}-D\'{i}az}
\affiliation{Departament de F\'isica Qu\`antica i Astrof\'isica, 
Facultat de F\'{\i}sica, Universitat de Barcelona, E--08028 Barcelona, Spain}
\affiliation{Institut de Ci\`encies del Cosmos, Universitat de 
Barcelona, ICCUB, Mart\'i i Franqu\`es 1, Barcelona 08028, Spain}
\affiliation{ICFO-Institut de Ciencies Fotoniques, The Barcelona Institute 
of Science and Technology, 08860 Castelldefels (Barcelona), Spain}

\date{\today}

\begin{abstract}
We investigate the leading area-law contribution to entanglement entropy in a system described by a general Lagrangian with O(2) symmetry containing first- and second-order time derivatives, namely breaking the Lorentz-invariance. We establish a connection between the Higgs gap present in a symmetry-broken phase and the area-law term for the entanglement entropy in the general, non-relativistic case. Our predictions for the entanglement entropy and correlation length are successfully compared to numerical results in two paradigmatic systems: the Mott insulator to superfluid transition for ultracold lattice bosons, and the ground state of ferrimagnetic systems.
\end{abstract}


\maketitle

In condensed-matter physics, relativistic quantum field theories often arise as low-energy effective approximations. However, in multiple situations the local Lorentz invariance is lost. As a prominent example, near a quantum phase transition~\cite{sachdev_2011}, a dynamical critical exponent different from one indicates the different scaling of space and time. A second example are non-Lorentz-invariant systems where the ground state spontaneously breaks a symmetry of the Hamiltonian. Goldstone's theorem ensures the presence of Nambu-Goldstone (NG) bosons at low energy~\cite{Goldstone1961,PhysRev.127.965}, but the lack of Lorentz invariance may dramatically change their dispersion relation~\cite{PhysRevLett.14.3,PhysRev.146.301,PhysRevD.49.3033,PhysRevX.4.031057}. Non-relativistic NG bosons have been extensively studied recently~\cite{PhysRevX.4.031057} and naturally appear in the many-body context~\cite{sym2020609}, e.g. in the presence of long-range interactions \cite{PhysRevB.95.245111}.

The nature of the low-energy excitations of a many-body system is deeply related to the quantum fluctuations in the ground state, and has a profund impact on the structure of quantum entanglement across the system. For instance, the scaling of ground-state entanglement entropy with the subsystem size displays a logarithmic violation of the so-called area law \cite{PhysRevLett.71.666,PhysRevD.34.373,CALLAN199455,Calabrese_2004,Casini_2009,RevModPhys.82.277} in one-dimensional gapless systems with short-range interactions \cite{Calabrese_2004}, or in the presence of a Fermi surface in any dimension~\cite{gioev-klich, wolf}; and in bosonic systems with spontaneous symmetry breaking, a subdominant universal additive logarithmic correction carries the nature of the Goldstone modes \cite{Metlitski-Grover, PhysRevB.95.245111}.

In the present paper we study the entanglement present in a general non-relativistic field-theoretical description with O(2) symmetry. The latter appears both in many condensed-matter-physics phenomena~\cite{sachdev_2011}, and in particle physics at finite chemical potential, e.g. kaon condensation \cite{SCHAFER200167} with an enlarged U(2) symmetry. We show that the dominant area-law prefactor of entanglement entropy acquires a universal contribution throughout the phase diagram, associated to the finite correlation length in the gapped disordered phase, and to an elusive ``Higgs correlation length'' in the gapless ordered phase, associated to amplitude fluctuations of the order parameter. We discuss the relevance of our analytical field-theory predictions for two prominent examples of many-body phenomena: the superfluid to Mott insulator quantum phase transition for ultracold lattice bosons; and ferrimagnets, which are gapless yet short-range-correlated systems.

{\bf Non-relativistic low-energy theory.} We consider the general O(2)-invariant
Lagrangian density ${\mathcal{L}}$ describing the dynamics of a complex field $\psi(\vec{r},t)=\left(\phi_1 +i \phi_2\right)/\sqrt{2}$ in $D+1$ dimensions:
\begin{equation}
\mathcal{L} / K = |(\partial_t-i\mu_r)\psi|^2 - c^2 |\nabla \psi|^2 - m^2|\psi|^2 - c_4 |\psi|^4 ~,
\label{Eq:Action2}
\end{equation}
where the global factor $K$ plays the role of a kinetic mass for the field degrees of freedom [see Eq.~\eqref{eq_full_hamiltonian}].
The (relativistic) chemical potential $\mu_r$ breaks the Lorentz invariance of the theory, and is relevant to many condensed-matter problems where Lorentz invariance is absent~\cite{PhysRevB.14.1165,sachdev_2011}. For certain systems, such as superconductors, the equations of motion have to be symmetric under complex conjugation as a consequence of particle-hole symmetry~\cite{doi:10.1146/annurev-conmatphys-031214-014350,Varma2002}. This imposes the coefficient of the first-order time derivative to vanish, and therefore $\mu_r=0$. On the contrary, in pure non-relativistic systems, the dynamics is driven by a Schr\"odinger equation and only contains first-order time derivatives, e.g. superfluid Helium. The Lagrangian Eq.~\eqref{Eq:Action2} was studied in the context of relativistic Bose-Einstein condensates~\cite{PhysRevD.24.426,PhysRevLett.66.683}, and more recently, in the study of non-relativistic NG bosons~\cite{PhysRevD.49.3033,PhysRevX.4.031057,Alvarez-Gaume2017}, 
which naturally appear in systems at finite chemical potential~\cite{PhysRevD.83.125009,PhysRevLett.110.011602,PhysRevLett.111.021601}, where the interplay between first- and second-order time derivatives  
can lead to the appearance of massive NG bosons~\cite{PhysRevD.90.085014}.

Our purpose is to study the ground-state bipartite entanglement entropy for systems effectively described, at low energy, by the Lagrangian Eq.~\eqref{Eq:Action2}. In particular, we shall consider the influence of the Lorentz-invariance-breaking chemical potential, especially at the quantum phase transition between the disordered phase and the long-range-ordered phase. We consider a subsystem $A$ immersed in an infinite ground state, and compute the von Neumann entropy $S$ of its reduced state $\rho_A = {\rm Tr}_{B} |\Psi_0 \rangle \langle \Psi_0|$, where ${\rm Tr}_B$ denotes the trace over $B$ degrees of freedom (the complement of $A$), and $|\Psi_0\rangle$ is the ground state. In $D=2, 3$, $S$ obeys a so-called area law over the whole phase diagram, namely it scales as $S = a{\cal A}$ up to subdominant corrections, where ${\cal A}$ is the area of the boundary between the $A$ and $B$ regions \cite{RevModPhys.82.277}. In gapless $D=1$ systems, $S$ may display a logarithmic violation of the area law, namely $S \propto \log L$ where $L$ is the length of $A$ subsystem \cite{Calabrese_2004}. Furthermore, throughout the paper, we shall not consider the subdominant corrections \cite{Metlitski-Grover}, and focus instead on the area-law coefficient $a$. Our main purpose is to identify universal contributions to $a$.

For our purpose, it is more convenient to work with the Hamiltonian formulation of the theory. We introduce the canonical moments $\pi_{1/2} = \frac{\delta {\cal L}}{\delta(\partial_t \phi_{1/2})} = K(\partial_t \phi_{1/2} \pm \mu_r \phi_{2/1})$. The Hamiltonian operator reads, in Fourier space:
\begin{equation}
\begin{split}
	\mathcal{H}({\bf{k}}) = &\frac{1}{2K}(\pi_1^2 + \pi_2^2) + \frac{K}{2}(m^2 + c^2 k^2)(\phi_1^2 + \phi_2^2) \\
	&+ \mu_r(\phi_1 \pi_2 - \phi_2 \pi_1) + \frac{K}{4}c_4(\phi_1^2 + \phi_2^2)^2 ~,
\end{split}
\label{eq_full_hamiltonian}
\end{equation}
where  $[\phi_a({\bf k}), \pi_b({\bf k'})]=i\delta_{a,b}\delta({\bf k} - {\bf k'})$, and $[\phi_a({\bf k}), \phi_b({\bf k'})] = [\pi_a({\bf k}), \pi_b({\bf k'})]=0$.
Throughout the paper, we shall consider a gaussian approximation to the ground state, accounting for harmonic quantum fluctuations around a saddle-point (mean-field) solution. Our focus is the entanglement content of such quantum fluctuations. At the mean-field level, two phases are found: a disordered phase for $m^2 \ge \mu_r^2$ (such that $\phi_1^{(0)} = \phi_2^{(0)} = \pi_1^{(0)} = \pi_2^{(0)} = 0$), and an ordered phase for $m^2 < \mu_r^2$, where the O(2) symmetry is spontaneously broken  (\textit{e.g.} $\phi_2^{(0)} = \pi_1^{(0)} = 0$, $\phi_1^{(0)} = \sqrt{(\mu_r^2 - m^2) / c_4}$, $\pi_2^{(0)} = -K\mu_r \phi_1^{(0)}$). The quadratic Hamiltonian governing the harmonic fluctuations around this mean-field solution is found upon replacing $\phi_{1/2} \rightarrow \phi_{1/2}^{(0)} + \phi_{1/2}$ and $\pi_{1/2} \rightarrow \pi_{1/2}^{(0)} + \pi_{1/2}$, and neglecting terms of order $O(\phi_{1/2}^3, \pi_{1/2}^3)$ and higher. We also subtract the mean-field ground-state energy contribution.

\begin{figure}
\centering
\includegraphics[width=1.\linewidth]{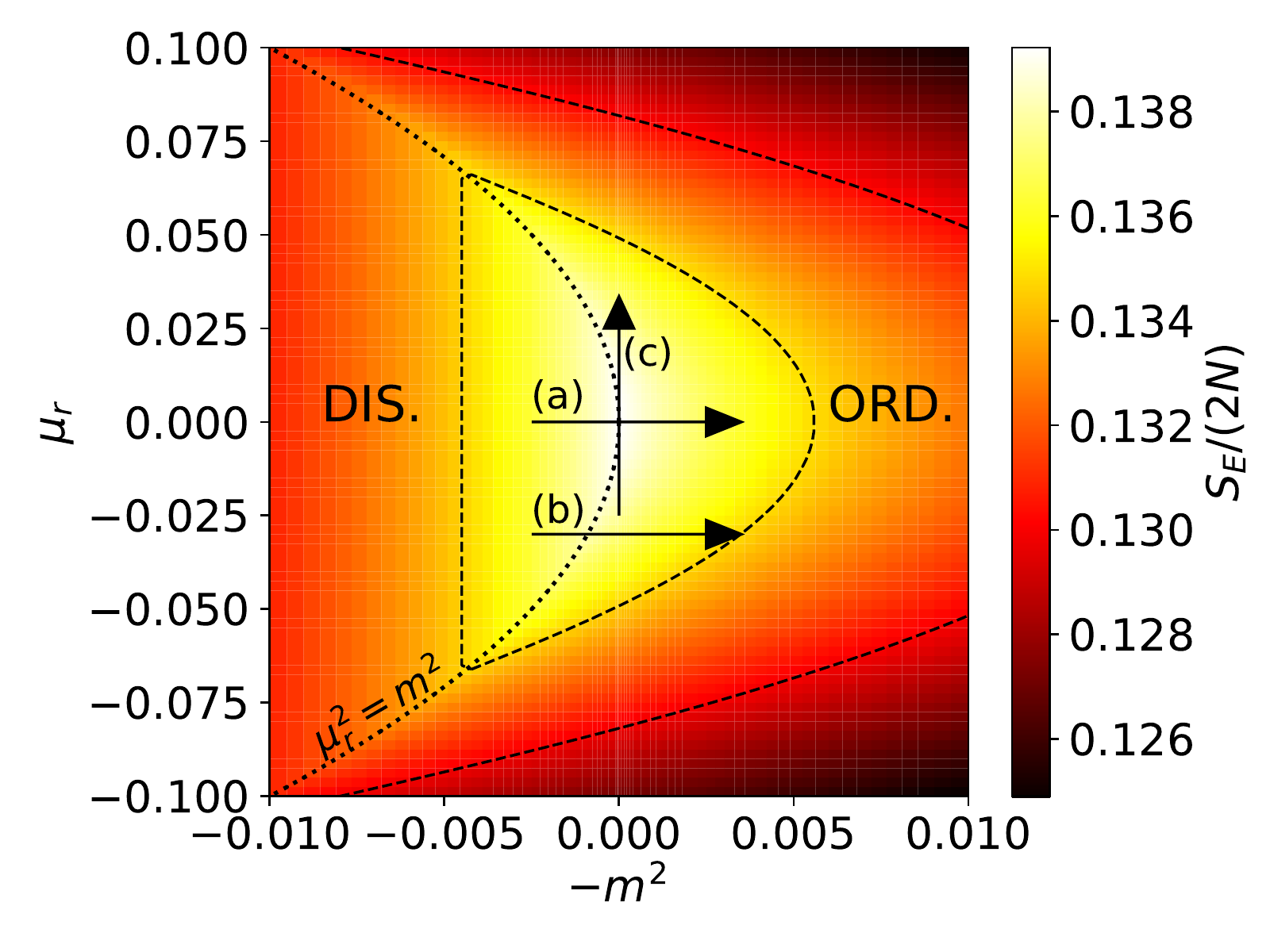}
\caption{Entanglement entropy per unit area across the 2D phase diagram of the model Eq.~\eqref{Eq:Action2}. Region $A$ is half of a $N \times N$ torus ($N=100$). Dotted line denotes the critical line $\mu_r^2=m^2$ separating the disordered (DIS.) and the ordered phase (ORD.). Dashed lines correspond to constant entanglement entropy. Arrows mark trajectories where the area-law prefactor is plotted on Fig.~\ref{fig:Entanglement2}.}
\label{fig:PhaseDiag}
\end{figure}

{\bf Disordered phase.} In the disordered phase ($m^2 \ge \mu_r^2$), the quadratic Hamiltonian is simply obtained from Eq.~\eqref{eq_full_hamiltonian} by setting $c_4 = 0$.
The excitation spectrum displays two gapped modes of engery $\omega_{\bf k}^{\pm} = \Omega_{\bf k} \pm \mu_r$, with $\Omega_{\bf k} = \sqrt{m^2 + c^2k^2}$ \cite{SuppMat}. Even though the Hamiltonian Eq.~\eqref{eq_full_hamiltonian} depends on $\mu_r$, its ground state is in fact independent of $\mu_r$. This is a manifestation of the Silver Blaze problem, \textit{i.e.} at zero temperature thermodynamical observables are independent of the chemical potential up to some critical value~\cite{PhysRevLett.91.222001}, namely $\mu_r^2 \le m^2$. Specifically, as shown in \cite{SuppMat}, we find a factorized ground-state wave-functional, $\Psi_0[\phi_1, \phi_2] = \Psi_{\Omega_{\bf k}}[\phi_1] \Psi_{\Omega_{\bf k}}[\phi_2]$, with:
\begin{equation}
	\Psi_{\Omega_{\bf k}}[\phi] \propto  \exp\left\{
		-\frac{1}{2} \int d^D {\bf k}~ K\Omega_{\bf k} \phi^2
	\right\} ~.
\end{equation}
This expression explicitly shows that quantum fluctuations in the ground-state are only sensitive to the combined excitation energy $\omega_{\bf k}^+ + \omega_{\bf k}^- = 2\Omega_{\bf k}$, as a consequence of the conservation of the charge associated to the O(2) invariance of the full theory. Furthermore, it shows that entanglement entropy $S$ is the sum of two contributions, stemming from the uncoupled gaussian fluctuations of the $\phi_1$ and $\phi_2$ fields. Entanglement entropy therefore obeys an area law containing a non-universal, UV-cutoff-dependent part $a_0$, and a universal contribution $a_1$ governed by the correlation length $\xi = c / m$ \cite{Metlitskietal2009, Calabrese_2004, SuppMat}:
\begin{equation}
	S / {\cal A} = a_0 - 2 a_1(\xi) ~,
	\label{eq_area_law_prefactors}
\end{equation}
where $a_1(\xi) = 1/(12\xi)$ in $D=2$ and $a_1(\xi) = (24 \pi \xi^2)^{-1} \log \xi$ in $D=3$.

{\bf Ordered phase.} In this part ($\mu_r^2 > m^2$), we focus on the ordered phase for systems in $D=2,3$ spatial dimensions, as in $D=1$ long-range order is not stable, and the physics is not captured by the gaussian approximation we consider. Here, $\phi_1$ and $\phi_2$ respectively capture amplitude and phase fluctuations of the order parameter $\psi^{(0)}=\sqrt{(\mu_r^2 - m^2) / c_4}$, and we find:
\begin{equation}
\begin{split}
	\mathcal{H}_{\rm ord.}^{(2)}({\bf k}) = \frac{1}{2K}(\pi_1^2 + \pi_2^2) + \mu_r(\phi_1 \pi_2 - \phi_2 \pi_1) \\
	+ \frac{K}{2} \left[
	\phi_1^2(c^2 k^2 + 3\mu_r^2 -2 m^2) + \phi_2^2(c^2 k^2 + \mu_r^2)
	\right] ~.
\end{split}
\label{eq_H2_ord}
\end{equation}
We first discuss the relativistic case ($\mu_r=0$, $m^2 < 0$). In this situation, amplitude and phase fluctuations are decoupled. They are the normal modes of the theory, namely the gapped Higgs mode with frequency $\omega^{\rm H}_{\bf k} = \sqrt{c^2 k^2 - 2 m^2}$ (the Higgs gap is $\Delta_{\rm H} = |m|\sqrt{2}$) and the gapless Goldstone mode with $\omega_{\bf k}^{\rm G} = ck$. As a consequence, the ground-state wave-functional factorizes: $\Psi_0[\phi_1, \phi_2] = \Psi_{\omega_{\bf k}^{\rm H}}[\phi_1] \Psi_{\omega_{\bf k}^{\rm G}}[\phi_2]$, so that entanglement entropy is again the sum of two contributions: $S = S_{\rm H} + S_{\rm G}$. Both contributions satisfy an area law in $D\ge 2$ spatial dimensions, and the area-law prefactor of $S_{\rm G}$ (stemming from the gapless Goldstone mode) is non-universal \cite{hastingsetal2010, humeniukR2012, Metlitski-Grover, PhysRevB.92.115129, swingle2016}. $S_{\rm H}$, on the other hand, contains a universal contribution $-a_1(\xi_{\rm H})$ to the area law prefactor \cite{Calabrese_2004}, governed by the ``Higgs correlation length'' $\xi_{\rm H} = c / \Delta_{\rm H}$. The expression of $a_1(\xi)$ is given after Eq.~\eqref{eq_area_law_prefactors}. This prediction is one of our main results. It shows that upon crossing the O(2) quantum-critical point, entanglement entropy displays a universal singularity, directly stemming from the Higgs mode going gapless at the critical point. Such a dependence on the correlation length was known in the (gapped) disordered phase \cite{Metlitskietal2009}, but not in the (gapless) ordered phase, where it is a consequence of the Higgs gap. This singularity has been observed in previous numerical studies \cite{Singhetal2012, HelmesW2014, PhysRevLett.116.190401}, and our paper provides its analytical explanation at a field-theory level.

In the general, non-relativistic case ($\mu_r \neq 0$), the normal modes (a gapped $\phi_{\rm H}$ and a gapless $\phi_{\rm G}$ mode) are linear combinations of $\phi_1$ and $\phi_2$ \cite{SuppMat}. The Goldstone mode maintains a linear dispersion at low energy, $\omega_{\bf k}^{\rm G} \approx c_{\rm G} k$, albeit with a modified velocity $c_{\rm G} = c\sqrt{(\mu_r^2 - m^2)/(3\mu_r^2 - m^2)}$. The Higgs gap is $\Delta_{\rm H} = \sqrt{6 \mu_r^2 - 2 m^2}$, and most importantly, it remains finite at the phase transition ($\Delta_H = 2|\mu_r|$). In this non-relativistic case, entanglement entropy cannot be separated into two additive contributions stemming from amplitude and phase fluctuations. We can, however, evaluate it numerically \cite{SuppMat}, following well-established methods to compute entanglement entropy for gaussian states \cite{Peschel_2009,PhysRevB.92.115129}. On Fig. \ref{fig:PhaseDiag}, we show entanglement entropy in the $(-m^2, \mu_r)$ phase diagram for $D=2$. Interestingly, the non-relativistic quantum ``critical'' points are actually not critical, as all correlation functions decay exponentially -- with a correlation length governed by the gap, $\xi=c/(2 |\mu_r|)$. This feature is clear from the discussion in the disordered phase: the ground state is independent of $\mu_r$ as long as $\mu_r \le m$, including at the ``critical'' point $\mu_r = m$. Therefore, it is identical to the ground state with $\mu_r=0$ and $m>0$, namely, strictly inside the disordered phase. This observation is key to understand the low-energy properties of ferrimagnets, as discussed at the end of the paper.

The results presented above are very general and affect a large variety of models 
whose low energy dynamics is captured by the Lagrangian Eq.~\eqref{Eq:Action2}. 
In the following we provide two prominent examples where our results can be directly 
applied.

{\bf The Bose-Hubbard model.} 
We first consider a paradigmatic instance of the O(2) quantum phase transition: the 
Mott-insulator (MI) to superfluid (SF) transition in the Bose-Hubbard model in $D=2$ dimensions \cite{Greiner2002}. The Hamiltonian describes a square-lattice Bose gas with contact interactions at zero temperature~\cite{PhysRevB.40.546}:
\begin{equation}
H=-J\sum_{\langle i,j\rangle} \left( b_i^{\dagger}b_j^{} + \text{h.c.} \right)
+\frac{U}{2}\sum_i b_i^{\dagger} b_i^{\dagger} b_i^{} b_i^{} - \mu \sum_i b_i^{\dagger}b_i^{},
\label{Eq:BH}
\end{equation}
where $b_i$ ($b_i^{\dagger}$) are bosonic annihilation (creation) operators 
on site $i=1,...,N$, $\mu$ is the chemical potential, $J$ is the hopping amplitude 
and $U$ is proportional to the two-bosons interaction strength. Near the critical point an effective low-energy description of the system applies~\cite{PhysRevB.40.546,PhysRevA.71.033629,sachdev_2011}. This effective description coincides with the Lagrangian Eq.~\eqref{Eq:Action2}, where 
the order parameter is proportional to the vacuum expectation value of the bosonic 
annihilation operator $\psi(\vec{r},t) \propto \langle b_i(t) \rangle$. 
The coefficients of the Lagrangian Eq.~\eqref{Eq:Action2} can be expressed 
in terms of the Bose-Hubbard parameters~\cite{sachdev_2011,PhysRevA.99.023614}, see Ref.~\cite{SuppMat}.
\begin{figure}[t]
\includegraphics[width=1.0\columnwidth]{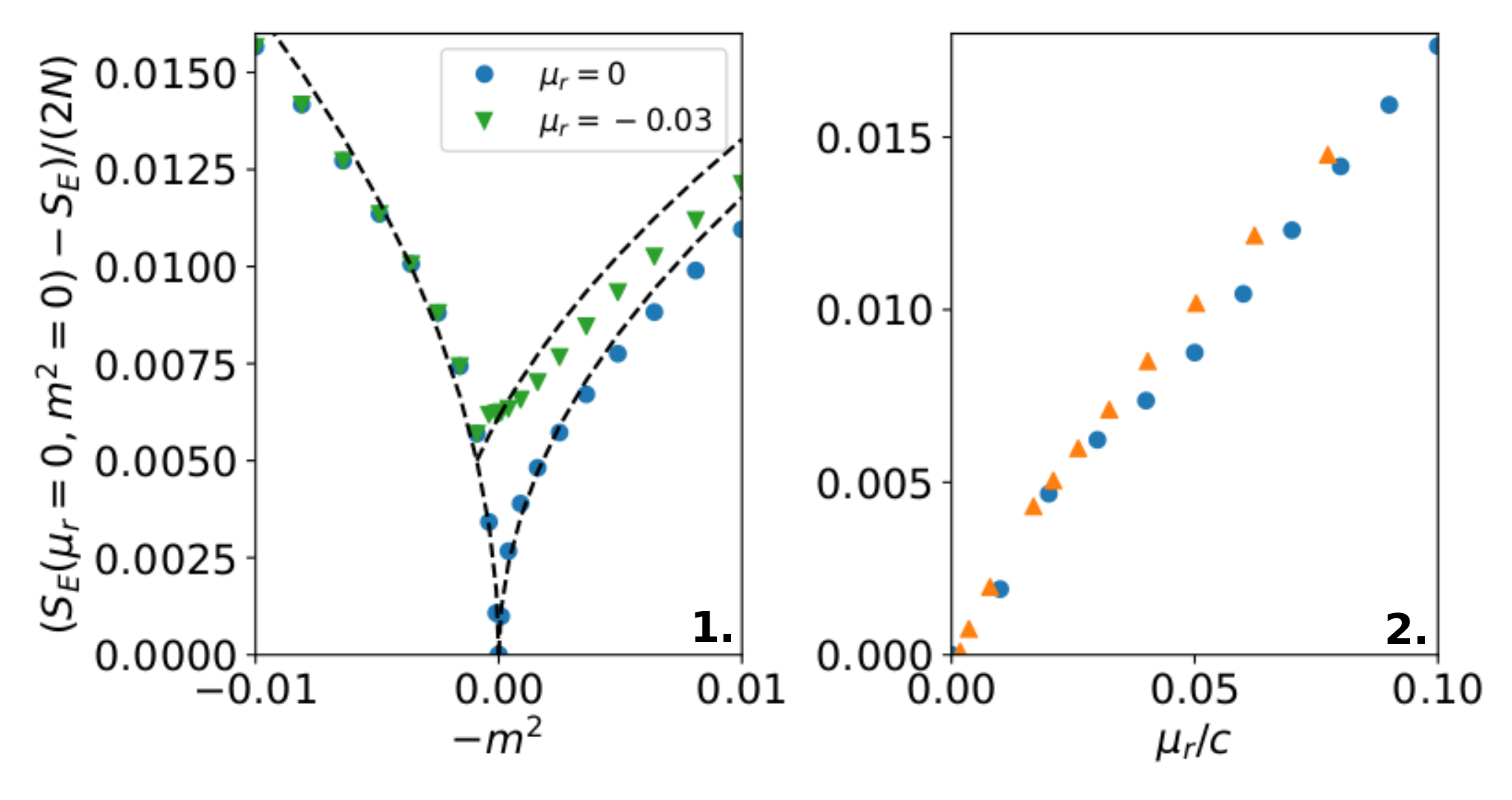}
\caption{Singular behaviour of the area-law coefficient in the thermodynamic limit. (1) Trajectories 
(a,b) of Fig.~\ref{fig:PhaseDiag}. Dashed lines correspond to Eq.~\eqref{eq_area_law_prefactors} using $\xi=c/m$ in the disordered phase and $\xi_H=c/\Delta_H$ in the ordered one. (2) Trajectory (c). 
Triangles: 2D Bose-Hubbard results (slave-bosons method of Ref.~\cite{PhysRevLett.116.190401}); circles: numerics on the gaussian field-theory \cite{SuppMat}.}
\label{fig:Entanglement2}
\end{figure}

The transition between the MI and the SF may occur via two different mechanisms: either the interaction strength is varied at fixed, integer filling fraction, or the particle-number is changed by adjusting the chemical potential. In the former case, the low-energy description becomes relativistic \cite{doi:10.1146/annurev-conmatphys-031214-014350} so that $\mu_r=0$ \cite{SuppMat}. The phase transition occurs across the O(2) quantum-critical point, where the Higgs correlation length $\xi_{\rm H}$ diverges. In the latter case, relativistic invariance is not present ($\mu_r \neq 0$), and the Higgs gap remains finite at the transition. The comparison between our results for the entanglement 
entropy along these two paths in the phase diagram, and the analytical formula Eq.~\eqref{eq_area_law_prefactors} using $\xi=c/m$ in the disordered phase and $\xi_{\rm H}=c/\Delta_{\rm H}$ in the ordered one, is shown on Fig.~\ref{fig:Entanglement2} (panel 1). The agreement is in all cases extremely good. Finally, on the SF side, touching the O(2) point by varying $\mu$ at fixed $U=U_c$, we predict that entanglement entropy behaves linearly with $\mu$, in agreement with the numerical results of Ref.~\cite{PhysRevLett.116.190401} [Fig.~\ref{fig:Entanglement2} (panel 2)].

{\bf Non-relativistic Nambu-Goldstone bosons}. 
The Lagrangian Eq.~\eqref{Eq:Action2} with $c_4=0$ and $\mu_r^2=m^2$ has 
been proposed as the low-energy description of NG bosons 
without Lorentz invariance~\cite{PhysRevD.49.3033,PhysRevX.4.031057}, in a physical situation where the system has a rotational O(3) symmetry which 
is broken down to O(2), \textit{i.e.} the ground state chooses a 
particular orientation. Following the general classification given in~\cite{NIELSEN1976445,PhysRevLett.108.251602,PhysRevLett.110.091601}, the original complex scalar field 
can be identified with two NG fields 
$\psi(\vec{r},t)=\pi_1(\vec{r},t)+i \pi_2(\vec{r},t)$. 
One of them corresponds to a type-B NG boson and has a quadratic dispersion 
relation~\cite{PhysRevX.4.031057}; the other one is a so-called gapped partner~\cite{PhysRevD.83.125009,PhysRevLett.110.011602,PhysRevLett.111.021601,PhysRevD.90.085014}.

\begin{figure}
\centering
\includegraphics[width=1.\columnwidth]{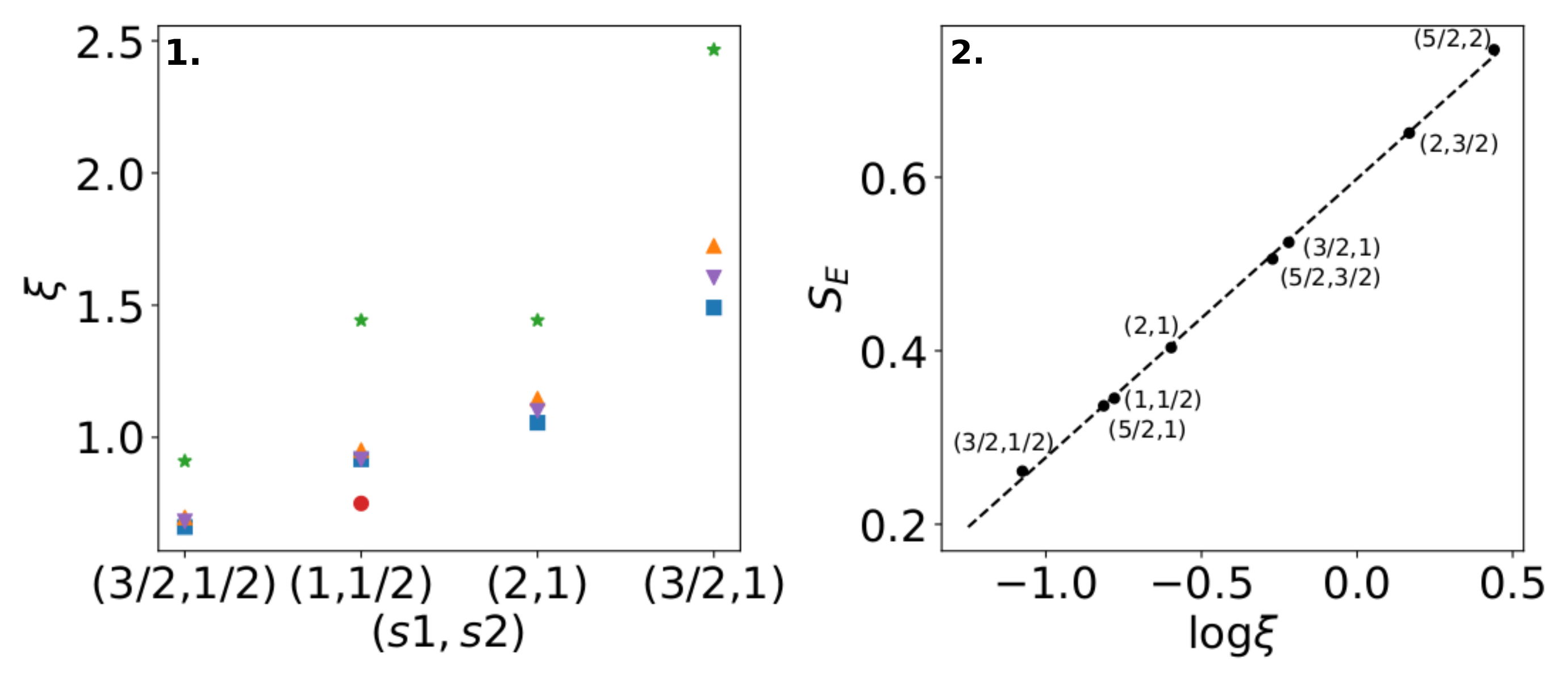}
\caption{Panel 1: Correlation lengths for different ferrimagnetic systems 
with spins $(s_1,s_2)$. Predicted values from Eq.~\eqref{Eq:CorrFerri} 
using previously computed values for $\rho$ and $\Delta$ from 
spin-wave~\cite{PhysRevB.55.8894} (stars), 
interacting spin wave~\cite{PhysRevB.55.8894} (upper triangles), and 
Monte-Carlo and exact diagonalization~\cite{PhysRevB.55.8894,Yamamoto2000} (squares). For $(s_1,s_2)=(1,1/2)$ we include the value obtained using matrix product states in~\cite{PhysRevB.55.8894,PhysRevB.55.R3336} (circles). Lower triangles are results obtained using iDMRG. Panel 2: Entanglement entropy for different ferrimagnets as a function of the extracted correlation length.}
\label{fig:CorrFerri}
\end{figure}

From our previous discussion, one sees that these systems are at the critical point between the disordered and the ordered phase, at finite chemical potential $\mu_r=m$. They exhibit a finite correlation 
length $\xi=c/m$ although they present 
quadratic gapless excitations. Therefore, entanglement entropy should obey an area 
law [see Eq.~\eqref{eq_area_law_prefactors}]. This should be compared with the 
relativistic case where (type-A) NG bosons acquire a linear dispersion 
relation, and a diverging correlation length.

Here, we focus on the $D=1$ case where entanglement 
entropy is expected to scale according to $S_E \sim \log \xi$~\cite{CALLAN199455,Calabrese_2004,PhysRevLett.106.050404}.
We validate our approach by considering a particular family of systems, namely ferrimagnets, which are spin systems living on two sublattices $A_1 \cup A_2$. On the $A_1$ sublattice, we define $\vec{S}_i$ as spin-$s_1$ operators, and on the $A_2$ sublattice, we define $\vec{\tau}_j$ as spin-$s_2$ operators. Spins interact via Heisenberg-type exchange interactions with Hamiltonian $H_{\rm ferrimag.}=J\sum_{<i,j>} \vec{S}_i \cdot \vec{\tau}_j$. Typically, the ground state of the system exhibits 
ferrimagnetic order, \textit{i.e.} an anti-alignment of the spins living 
on different sublattices. In this case, the different magnitude of 
the spins induces a total magnetization density in the system 
$m = \langle S^z \rangle / N \sim |s_1-s_2|$.  
This is an indicator of the non-zero expectation value of 
the commutator of two broken charges, $\langle\left[ S^x,S^y \right] \rangle\neq 0$, which breaks Lorentz invariance~\cite{PhysRevX.4.031057}.

The low-energy effective description of ferrimagnets is the Lagrangian Eq.~\eqref{Eq:Action2} with $c_4=0$ and $\mu_r^2 = m^2$ ~\cite{10.1093}, whose low-energy excitations are $\omega_{\bf k}^{\pm} = \sqrt{m^2 + c^2 k^2} \pm m$. Thus, we expect quadratic type-B NG bosons $\omega_{\bf k}^- = \rho k^2$ and a gapped partner $\omega_{\bf k}^+ = \Delta$, where $\rho$ and $\Delta$ are the spin stiffness and energy gap, respectively. In terms of the coefficients of the Lagrangian Eq.~\eqref{Eq:Action2}, we identify $\rho=c^2/(2m)$ and $\Delta=2m$, which allows us to write the expression 
for the correlation length (in units of the lattice spacing):
\begin{equation}
\xi=2\sqrt{\frac{\rho}{\Delta}} \,.
\label{Eq:CorrFerri}
\end{equation}

We compare the prediction Eq.~\eqref{Eq:CorrFerri} for the correlation length with various numerical computations on $H_{\rm ferrimag.}$, where a very short correlation length was found together with gapless excitations~\cite{PhysRevB.55.8894,PhysRevB.55.R3336,PhysRevB.57.13610}, two key features which are 
clearly present at the field-theory level. Indeed, Eq.~\eqref{Eq:CorrFerri} allows 
us to predict the value of the correlation length for several ferrimagnets
using previously obtained numerical results for $\rho$ and $\Delta$. 
The results are summarized on Fig.~\ref{fig:CorrFerri}. Our prediction for $\xi$ is systematically smaller than those predicted by spin-wave theory, $\xi^{-1}=\log(s_1/s_2)$~\cite{PhysRevB.55.8894}, and closer to the (more accurate) value obtained using matrix product states~\cite{PhysRevB.55.8894,PhysRevB.55.R3336} for $(s_1=1,s_2=1/2)$. In addition to existing results in the literature, we also carried out iDMRG simulations~\cite{tenpy} and computed the correlation lengths for different spin values $s_1$ and $s_2$. We found a good agreement with the prediction Eq.~\eqref{Eq:CorrFerri}, using existing Monte-Carlo data for the values of $\rho$ and $\Delta$. Finally, we observe that these values of $\xi$ are also consistent with entanglement entropy scaling as $S \sim \log \xi$.

{\bf Outlook} We have investigated the area-law prefactor of entanglement entropy in non-relativistic low-energy field theories with O(2) symmetry. Our predictions have been successfully confronted with two prominent examples from condensed-matter physics: the Mott insulator to superfluid transition and non-relativistic Nambu-Goldstone bosons in ferrimagnets. Our findings, which could be tested in quantum simulators~\cite{Islam2015, Brydges260}, also apply to particle-physics models with a non-zero chemical potential.

\begin{acknowledgments} 
The authors thank Joan Martorell for a careful reading of the manuscript. The authors also thank Tomeu Fiol, Luca Tagliacozzo, Josep Taron, and Germ\'an Sierra for useful comments and discussions. 
This work is partially funded by MINECO (Spain) (Grant No. FIS2017-87534-P and Severo Ochoa SEV-2015-0522), the
European Union Regional Development Fund within the
ERDF Operational Program of Catalunya (project QUASICAT/QuantumCat), the fundacio Mir-Puig and Cellex through an ICFO-MPQ
postdoctoral fellowship, and the Generalitat de Catalunya (SGR 1381 and CERCA Programme).
\end{acknowledgments}

\bibliographystyle{apsrev4-1}
\bibliography{paperbib}

\onecolumngrid
\clearpage
\appendix
\section*{SUPPLEMENTAL MATERIAL for \textit{Entanglement entropy in low-energy field theories at finite chemical potential}}

In this Supplemental Material we provide further details about:

1. The representation of the ground state wavefunctional in real in the disordered phase and its connection with entanglement entropy.

2. The gaussian approximation in the ordered phase and the computation of entanglement entropy.

3. The explicit relation with the parameters of the Bose-Hubbard model.

\subsection{1. Disordered phase}
We consider the free non-relativistic theory describing the dynamics of a complex scalar field $\psi = \left(\phi_1 +i \phi_2\right)/\sqrt{2}$, whose Lagrangian density in $D+1$-dimensional spacetime is given by,
\begin{equation}
	{\cal L} / K = |(\partial_t - i\mu_r)\psi|^2 - c^2|\nabla \psi|^2 - m^2|\psi|^2 - c_4|\psi|^4 ~.
\label{Eq:action}
\end{equation}
The theory is non-relativistic because of the presence of the term $\mu_r \neq 0$. We work with the Hamiltonian formulation of the theory. We introduce the canonical moments $\pi_{1/2} = \frac{\delta {\cal L}}{\delta(\partial_t \phi_{1/2})} = K(\partial_t \phi_{1/2} \pm \mu_r \phi_{2/1})$. The Hamiltonian operator reads, in Fourier space:
\begin{equation}
	\mathcal{H}({\bf{k}}) = \frac{1}{2K}(\pi_1^2 + \pi_2^2) + \frac{K}{2}(m^2 + c^2 k^2)(\phi_1^2 + \phi_2^2)+ \mu_r(\phi_1 \pi_2 - \phi_2 \pi_1)~,
\end{equation}
where  $[\phi_a({\bf k}), \pi_b({\bf k'})]=i\delta_{a,b}\delta({\bf k} - {\bf k'})$, and $[\phi_a({\bf k}), \phi_b({\bf k'})] = [\pi_a({\bf k}), \pi_b({\bf k'})]=0$.

We are now neglecting the interactions, setting $c_4=0$. Therefore, the results derived in this section can be seen as performing a Gaussian approximation in the disordered phase described in the main text. 
We look for a canonical transformation of the form:
\begin{equation}
\begin{split}
\pi_{\pm}& = \frac{1}{\sqrt{2}} \left[
	 \sqrt{\frac{\Gamma_{\pm}}{K\Omega}} \pi_1 \pm 
	 \sqrt{K\Omega \Gamma_{\pm}}\phi_2
	 \right] ~,
\\
\phi_{\pm}&= \frac{1}{\sqrt{2}} \left[
	\sqrt{\frac{K\Omega}{\Gamma_{\pm}}} \phi_1 \mp 
	\frac{1}{\sqrt{K\Omega \Gamma_{\pm}}}\pi_2
	\right] ~.
\end{split}
\label{Eq:CanTrans}
\end{equation}
One can verify that choosing $\Gamma_{\pm} = \sqrt{m^2 + c^2 k^2} \mp \mu_r$ and $\Omega = \sqrt{m^2 + c^2 k^2}$, the Hamiltonian reads:
\begin{equation}
\mathcal{H}({\bf{k}})=\frac{1}{2}\left( \pi_+^2 + \pi_-^2 \right)+\frac{1}{2} (\Gamma_+^2 \phi_+^2 + \Gamma_-^2 \phi_-^2) ~.
\label{Eq:Ham2}
\end{equation}
Finally the Hamiltonian Eq.~\eqref{Eq:Ham2} can be diagonalized by introducing a set of normal modes ($\pm$) defined by the annihilation operators:
\begin{equation}
a_{\pm}=\sqrt{\frac{\Gamma_{\pm}}{2}}\phi_{\pm}+ \frac{i}{\sqrt{2 \Gamma_{\pm}}} \pi_{\pm} ~,
\label{Eq:a}
\end{equation}
which allows one to write the Hamiltonian in an harmonic-oscillator form,
\begin{equation}
\mathcal{H}({\bf{k}}) = \Gamma_+ \left( a_+^{\dagger}a_+ +  \frac{1}{2} \right) +  \Gamma_- \left( a_-^{\dagger}a_- +  \frac{1}{2} \right).
\end{equation}
Because of this harmonic-oscillator form we know that the ground state $|\Psi_0\rangle$ is uniquely determined by the condition $a_+({\bf{k}})|\Psi_0\rangle = a_-({\bf{k}})|\Psi_0\rangle = 0$, $\forall ({\bf{k}})$. In the Schr\"odinger representation, the groundstate wavefunctional is the gaussian:
\begin{equation}
\Psi_0 \left[\phi_+,\phi_- \right]= \mathcal{N} e^{ -\frac{1}{2} \int d^D{\bf{k}} \left[ \phi_+ \Gamma_+ \phi_+ + \phi_- \Gamma_- \phi_- \right]},
\label{Eq:WaveFunc}
\end{equation}
where $\mathcal{N}$ is a normalization factor. Turning back to the original $\phi_1$ and $\phi_2$ fields using the canonical transformation Eq.~\eqref{Eq:CanTrans}, one obtains:
\begin{equation}
\Psi_0 \left[\phi_1, \pi_2 \right]= \mathcal{N} e^{ -\frac{1}{2} \int d^D{\bf{k}} \left[ \phi_1K \Omega \phi_1 + \pi_2 \frac{1}{K\Omega} \pi_2 \right]} ~,
\end{equation}
or equivalently:
\begin{equation}
\Psi_0 \left[\phi_1, \phi_2 \right]= \mathcal{N} e^{ -\frac{1}{2} \int d^D{\bf{k}} \left[ \phi_1 K \Omega \phi_1 + \phi_2 K \Omega \phi_2 \right]} ~,
\end{equation}
which coincides with the groundstate wavefunctional of two relativistic scalar fields of kinetic mass $K$ and dispersion relation $\Omega({\bf k}) = \sqrt{m^2 + c^2 k^2}$. This result explicitly shows that the ground state is independent of $\mu_r$, as long as we stay in the disordered phase (namely, as long as $\mu_r^2 \le m^2$).
From this groundstate wavefunctional one can readily see that the two-point function $\langle0| \psi(\vec{x},t) \psi^{\dagger}(\vec{y},t)|0\rangle$ has an exponential decay $\exp\{ -r/\xi \}$ for large distance $r=|\vec{x}-\vec{y}|$ given by the correlation length $\xi = c/m$.
Since we have been able to express the groundstate wavefunctional as the one of a relativistic complex scalar field with a mass $m$, the computation of entanglement entropy will be based on the known results obtained for the relativistic case~\cite{PhysRevLett.71.666,PhysRevD.34.373,CALLAN199455,Calabrese_2004,Casini_2009}.

Consider that we perform a spatial partition of the system into two parts $\mathcal{A}$ and $\mathcal{B}$. The entanglement entropy is defined as $S_E=-\text{Tr}_{\mathcal{A}} \{\rho_{\mathcal{A}} \ln \rho_{\mathcal{A}}  \}$, where $\rho_{\mathcal{A}} = \text{Tr}_{\mathcal{B}} \rho$ is the reduced density matrix obtained by tracing out the degrees of freedom living on $\mathcal{B}$ and $\rho=|0\rangle \langle 0|$ is the density matrix which can be represented using the groundstate wavefunctional Eq.~\eqref{Eq:WaveFunc}.
If the system presents a finite correlation length $\xi$ the leading area law term of entanglement entropy is given by~\cite{Calabrese_2004,PhysRevLett.106.050404},
\begin{equation}
S_E=-A\frac{N}{12}\int \frac{d^{D-1}k_{\perp}}{(2\pi)^{D-1}}\log \frac{k_{\perp}^2 + \xi^{-2}}{k_{\perp}^2+a^{-2}},
\label{Eq:ENT}
\end{equation} 
where $\epsilon=1/a$ is the UV-cutoff, $A$ is the area of the boundary separating the two parts of the system $\mathcal{A}$ and $\mathcal{B}$, and $N=2$ is the number of free scalar fields. One can see that this integral contains a term which is UV-cutoff dependent and that makes it non-universal. On the other hand, there is a finite universal term that goes like $\xi^{1-D}$ which can be explicitly manifested by taking the derivative $\partial S_E/\partial \xi^{-2}$,
\begin{equation}
\frac{\partial S_E}{\partial \xi^{-2}} = -A\frac{N}{12}\int \frac{d^{D-1}k_{\perp}}{(2\pi)^{D-1}}\frac{1}{k_{\perp}^2 + \xi^{-2}},
\end{equation}
which leads to the well known result for $D=2$~\cite{Calabrese_2004,PhysRevLett.106.050404},
\begin{equation}
S_E/A = -\frac{1}{6\xi} + a_0,
\label{Eq:ENT2}
\end{equation}
where $a_0$ stands for some non-universal divergent constant cutoff dependent. In this way one can compute $\Delta S_E=S_E(\xi\rightarrow\infty)- S_E(\xi)$ which should not depend on the UV cutoff and therefore it will be universal in this sense.

\subsection{2. Normal modes in the ordered phase}
We consider the Lagrangian density:
\begin{equation}
	{\cal L} / K = |(\partial_t - i\mu_r)\psi|^2 - c^2|\nabla \psi|^2 - m^2|\psi|^2 - c_4|\psi|^4 ~.
\end{equation}
Introducing $\psi = (\phi_1 + i\phi_2) / \sqrt{2}$, and $\pi_{1/2} = \frac{\partial{\cal L}}{\partial(\partial_t \phi_{1/2})} = K(\partial_t \phi_{1/2} \pm \mu_r \phi_{2/1})$, we have the equivalent Hamiltonian density:
\begin{eqnarray}
	{\cal H} & = & \pi_1 \partial_t \phi_1 + \pi_2 \partial_t \phi_2 - {\cal L} \\
	&=& \frac{1}{2K}(\pi_1^2 + \pi_2^2) + \mu_r(\pi_2 \phi_1 - \pi_1 \phi_2) + \frac{Km^2}{2}(\phi_1^2 + \phi_2^2) + \frac{Kc^2}{2}[(\nabla \phi_1)^2 + (\nabla \phi_2)^2] + \frac{Kc_4}{4}(\phi_1^2 + \phi_2^2)^2 ~.
\end{eqnarray}
We focus on the ordered phase, namely $m^2 < \mu_r^2$. Imposing that $\delta {\cal H}=0$, we obtain the mean-field solution $\phi_2^{(0)}=0$, $\pi_1^{(0)}=0$, $\pi_2^{(0)}=-K \mu_r \phi_1^{(0)}$, and $\phi_1^{(0)} = \pm\sqrt{(\mu_r^2 - m^2) / c_4}$. Expanding around this mean-field solution (namely, $x \rightarrow x^{(0)} + x$ for $x=\pi_{1/2}, \phi_{1/2}$) and neglecting terms of order $O(x^3)$, we obtain: 
\begin{equation}
	{\cal H}^{(2)} = \frac{1}{2K}(\pi_1^2 + \pi_2^2) + \mu_r(\pi_2 \phi_1 - \pi_1 \phi_2) + \frac{K}{2}\mu_r^2(\phi_1^2 + \phi_2^2) + K(\mu_r^2 - m^2)\phi_1^2 + \frac{Kc^2}{2}[(\nabla \phi_1)^2 + (\nabla \phi_2)^2] ~.
\end{equation}
Clearly, when $\mu_r = 0$, $\phi_1$ and $\phi_2$ are decoupled and represent the normal modes (namely, the Higgs and Goldstone modes of the ordered phase). They contribute independently to entanglement entropy of a subsystem, which therefore reads $S = S_{\rm Goldstone} + S_{\rm Higgs}$. However, for $\mu_r \neq 0$ they are coupled. 
\subsubsection{Diagonalization of the Hamiltonian}
Going to Fourier space, the Hamiltonian density reads:
\begin{equation}
	 {\cal H}^{(2)}(k) = \frac{1}{2}\left(
	 	\begin{array}{cccc}
	 		\phi_1 & \phi_2 & \pi_1 & \pi_2
	 	\end{array}
	 	\right) H(k) \left(
	 	\begin{array}{c}
	 		\phi_1 \\ \phi_2 \\ \pi_1 \\ \pi_2
		\end{array}	 	 
		\right) ~,
\end{equation} 
where $H$ is of the form: 
\begin{equation}
	H(k) = \left(\begin{array}{cccc}
		A_1 & 0 & 0 & C \\
		0 & A_2 & -C' & 0 \\
		0 & -C' & B_1 & 0 \\
		C & 0 & 0 & B_2
	\end{array} \right) ~,
\end{equation}
with $A_1=K(3 \mu_r^2 - 2m^2 + c^2k^2)$, $A_2=K(\mu_r^2 + c^2 k^2)$, $B_1=B_2 = 1/K$, $C=C'=\mu_r$.
In order to diagonalize the Hamiltonian, we look for a canonical transformation
$
	\left(
	 	\begin{array}{c}
	 		\phi_1 \\ \phi_2 \\ \pi_1\\ \pi_2
	 	\end{array}
	 	\right) = W \left(
	 	\begin{array}{c}
	 		q_1 \\ q_2 \\ p_1 \\ p_2
	 	\end{array}
	 	\right)
$, where $W$ is chosen of the form: $
	W = \left(\begin{array}{cccc}
		\alpha_1 & \alpha_2 & 0 & 0 \\
		0 & 0 & \beta_1 & \beta_2 \\
		0 & 0 & \gamma_1 & \gamma_2 \\
		\delta_1 & \delta_2 & 0 & 0 \\	
	\end{array}\right)
$.
$W$ represents a canonical transformation iff $WJW^T=J$ with $J=\left(\begin{array}{cccc}0 & 0 & 1 & 0 \\ 0& 0 & 0 & 1 \\ -1&0&0&0 \\ 0&-1&0&0 \end{array}\right)$. Introducing the matrices $X = \left(\begin{array}{cc}\alpha_1 & \alpha_2 \\ \delta_1 & \delta_2 \end{array}\right)$ and $Y = \left(\begin{array}{cc}\gamma_1 & \gamma_2 \\ -\beta_1 & -\beta_2 \end{array}\right)$, $W$ is a canonical transformation iff $XY^T = {\rm Id}$. The aim of the canonical transformation $W$ is to bring the matrix $H$ into the form: 
\begin{equation}
	W^T H W = \left(\begin{array}{cccc}
		\omega_1^2 & 0 & 0 & 0 \\
		0 & \omega_2^2 & 0 & 0 \\
		0 & 0 & 1 & 0 \\
		0 & 0 & 0 & 1 \\	
	\end{array}\right) ~.
\end{equation}
This is equivalent to having $X^T M_1 X = \left(\begin{array}{cc}\omega_1^2 & 0 \\ 0 & \omega_2^2 \end{array}\right)$ with $M_1 = \left(\begin{array}{cc}A_1 & C \\ C & B_2 \end{array}\right)$; and $Y^T M_2 Y = {\rm Id}$ with $M_2 = \left(\begin{array}{cc}B_1 & C' \\ C' & A_2 \end{array}\right)$. The diagonalization may be achieved by noticing that $\left(\begin{array}{cc}\omega_1^2 & 0 \\ 0 & \omega_2^2 \end{array}\right) = X^T M_1 X Y^T M_2 Y = X^T M_1 M_2 Y$, where we use the fact that $XY^T={\rm Id}$. Since $X^T = Y^{-1}$, we can now diagonalize $M_1 M_2$ to obtain $X$, $Y$, $\omega_1$ and $\omega_2$. This is done in two steps. First, $M_1 M_2$ is diagonalized, yielding $\tilde{X}^T M_1 M_2 \tilde{Y} = {\rm diag}(\omega_1^2, \omega_2^2)$. $Y$ is then obtained by properly normalizing the columns of $\tilde{Y}$. We initially have $\tilde{Y}^T M_2 \tilde{Y} = {\rm diag}(\lambda_1^2, \lambda_2^2)$. Finally, we obtain the normal modes by choosing $Y=\tilde{Y}{\rm diag}(1/\lambda_1, 1 / \lambda_2)$, and $X^T=Y^{-1}$. Then the Hamiltonian reads:
\begin{equation}
	{\cal H}^{(2)}(k) = \sum_{i=1}^2 \frac{1}{2}\left[p_i(k)p_i(-k) + \omega_i(k)^2 q_i(k)q_i(-k)\right] ~,
\end{equation}
where $\left(\begin{array}{c}
	\phi_1 \\ \pi_2
\end{array} \right) = X \left(\begin{array}{c}
	q_1 \\ q_2
\end{array}\right)$ and $\left(\begin{array}{c}
	\pi_1 \\ -\phi_2
\end{array}\right) = Y\left(\begin{array}{c}
	p_1 \\ p_2
\end{array}\right)$. 

\subsubsection{Eigenmodes} 
The eigenfrequencies are:
\begin{equation}
	\omega_{\pm}^2 = c^2 k^2 + 3 \mu_r^2 - m^2 \pm \sqrt{
		4\mu_r^2 c^2 k^2 + (3 \mu_r^2 - m^2)^2
		} ~.
\end{equation}
For the gapless Goldstone mode, we find, expanding at low $k$:
\begin{equation}
	\omega_{\rm G}^2 = \omega_-^2 = c^2 k^2 \frac{\mu_r^2 - m^2}{3 \mu_r^2 - m^2} + O(k^4) ~,
\end{equation}
Hence a modified sound velocity $c_{\rm G} = c \sqrt{\frac{\mu_r^2 - m^2}{3 \mu_r^2 - m^2}}$, such that $\omega_{\rm G} = c_{\rm G} k + O(k^2)$. For the gapped Higgs mode, we find instead:
\begin{equation}
	\omega_{\rm H}^2 = \omega_+^2 = 6\mu_r^2 - 2m^2 + c^2 k^2 \frac{5\mu_r^2 - m^2}{3 \mu_r^2 - m^2} + O(k^4) ~.
\end{equation}
This defines the Higgs gap: $\Delta_{\rm H} = \sqrt{6\mu_r^2 - 2m^2}$ and the Higgs velocity: $c_{\rm H} = 
c \sqrt{\frac{5\mu_r^2 - m^2}{3 \mu_r^2 - m^2}}$, such that $\omega_{\rm H} = \sqrt{\Delta_{\rm H}^2 + c_{\rm H}^2 k^2} + O(k^4)$.

\subsubsection{Correlation functions}
Correlation functions in the ground state are obtained according to:
\begin{equation}
	\left\langle \left( \begin{array}{c}
		\phi_1 \\ \pi_2
	\end{array} \right) \left( \begin{array}{cc}
		\phi_1 & \pi_2
	\end{array} \right)
	\right\rangle = X \left(\begin{array}{cc}
		\frac{1}{2\omega_1}& 0 \\ 0& \frac{1}{2\omega_2}
	\end{array} \right) X^T
\end{equation}
\begin{equation}
	\left\langle \left( \begin{array}{c}
		\pi_1 \\ -\phi_2
	\end{array} \right) \left( \begin{array}{cc}
		\pi_1 & -\phi_2
	\end{array} \right)
	\right\rangle = Y \left(\begin{array}{cc}
		\omega_1/2& 0 \\ 0& \omega_2/2
	\end{array} \right) Y^T = \frac{1}{4}\left\langle \left( \begin{array}{c}
		\phi_1 \\ \pi_2
	\end{array} \right) \left( \begin{array}{cc}
		\phi_1 & \pi_2
	\end{array} \right)
	\right\rangle^{-1} ~,
\end{equation}
and $\langle \phi_1 \phi_2 \rangle = 0 = \langle \pi_1 \pi_2 \rangle$, $\langle \phi_1 \pi_1 \rangle = i/2 = \langle \phi_2 \pi_2 \rangle$. 

\subsubsection{Entanglement entropy}
In order to compute the entanglement entropy of a subsystem $A$, one first forms the correlation matrix for the field degrees of freedom belonging to that subsystem: $\langle \phi_1(x) \phi_2(x') \rangle$ for $x, x' \in A$, etc. We assume that $A$ contains $N$ sites. The correlation matrix is:
\begin{equation}
	{\cal C} = \left(\begin{array}{cccc}
		\langle \phi_1 \phi_1\rangle & 0 & (i/2) {\rm Id} & \langle \phi_1 \pi_2\rangle \\
		0 & \langle \phi_2 \phi_2\rangle & \langle \phi_2 \pi_1\rangle & (i/2) {\rm Id} \\
		(-i/2) {\rm Id} & \langle \pi_1 \phi_2\rangle & \langle \pi_1 \pi_1\rangle & 0 \\
		\langle \pi_2 \phi_1\rangle & (-i/2){\rm Id} & 0 & \langle \pi_2 \pi_2\rangle
	\end{array} \right) ~,
\end{equation}
where $\langle \phi_1 \phi_1 \rangle$ represents the $N\times N$ matrix $\langle \phi_1(x) \phi_2(x') \rangle$ for $x, x' \in A$, and similarly for $\langle \phi_1 \pi_2 \rangle$, etc. Similarly to the diagonalization of the Hamiltonian, the canonical transformation to the normal modes may be chosen of the form: $W = \left(\begin{array}{cccc}
		\alpha_1 & \alpha_2 & 0 & 0 \\
		0 & 0 & \beta_1 & \beta_2 \\
		0 & 0 & \gamma_1 & \gamma_2 \\
		\delta_1 & \delta_2 & 0 & 0 \\	
	\end{array}\right)
$, where now $\alpha_i, \beta_i, \gamma_i, \delta_i$ are $N \times N$ matrices. The diagonalization of ${\cal C}$ goes along the same line as that of the Hamiltonian, except for the fact that each symbol now represents a $N \times N$ matrix. Indeed, introducing $X=\left(\begin{array}{cc}\alpha_1 & \alpha_2 \\ \delta_1 & \delta_2 \end{array} \right)$ and $Y=\left(\begin{array}{cc}\gamma_1 & \gamma_2 \\ -\beta_1 & -\beta_2 \end{array} \right)$, and denoting ${\cal C}= \left(\begin{array}{cccc}
		A_1 & 0 & (i/2){\rm Id} & C \\
		0 & A_2 & -C' & (i/2){\rm Id} \\
		-(i/2){\rm Id} & -C'^T & B_1 & 0 \\
		C^T & -(i/2){\rm Id} & 0 & B_2
	\end{array} \right)$, we have that ${\cal C} = {\rm diag}(X^TM_1X, ~Y^TM_2Y)$ with $M_1 = \left(\begin{array}{cc}A_1 & C \\ C^T & B_2 \end{array}\right)$ and $M_2 = \left(\begin{array}{cc}B_1 & C'^T \\ C' & A_2 \end{array}\right)$. Diagonalizing the matrix $M_1 M_2$, one obtains eigenvalues of the form $\lambda_i^2 = \left[\frac{1}{2} + \frac{1}{e^{\tilde{\omega}_i} - 1} \right]^2$, where $\tilde{\omega}_i$ forms the (one-body) entanglement spectrum. The entanglement entropy of $A$ is finally obtained as:
	\begin{equation}
		S = \sum_i \left[(\lambda_i + \frac{1}{2}) \log(\lambda_i + \frac{1}{2}) - (\lambda_i - \frac{1}{2})\log(\lambda_i - \frac{1}{2}) \right] ~.
	\end{equation}
	As a further simplification, we notice that if $A$ is periodic along a certain direction (say $x$), the correlation matrix ${\cal C}$ is block-diagonal with respect to the momentum $k_x$ along that direction. The diagonalization may thus be achieved separately in each momentum sector $k_x$, and the entanglement entropy is the sum of the contributions from the different sectors $k_x$.

\subsection{3. Relation with Bose-Hubbard parameters}
In this section we write down explicitly the relations between the Bose-Hubbard (BH) parameters and the effective parameters of our original model. We introduce the notations $K_1 \equiv 2\mu_r K$, $c_2\equiv K\left(m^2-\mu_r^2 \right)$ and $K_3\equiv K c^2 $. For the first lobe $n=1$ and setting $U=1$ the following relations can be applied~\cite{sachdev_2011,PhysRevA.99.023614},
\begin{equation}
c_2=\mu \frac{1-\mu }{1+\mu}-zJ, \quad K_1=-\frac{\partial c_2}{\partial \mu}=\frac{\mu^2 +2\mu -1}{(1+\mu)^2}, \quad K=-\frac{1}{2}\frac{\partial^2c_2}{\partial \mu^2}=\frac{2}{(1+\mu)^3},\quad K_3=J.
\end{equation}
In these relations the chemical potential of the BH model $\mu$ can be different from the chemical potential of the original model $\mu_r$ (we add the subscript $r$ to differentiate them). In order to simplify these relations we expand around the tip of the lobe $zJ_c=3-2\sqrt{2}$ and $\mu_c=\sqrt{2}-1$, and we obtain the following relations,
\begin{equation}
c_2/K \approx z \sqrt{2}(J_c-J)- (\mu-\mu_c)^2, \quad K_1\approx 2K(\mu_c)(\mu-\mu_c),
\label{Eq:BHrel2}
\end{equation}
where $K(\mu_c)=1/\sqrt{2}$.
If we compare these relations with the ones for the original model $c_2/K=m^2- \mu_r^2$ and $\mu_r=K_1/2K$, we see that at leading order,
\begin{equation}
m^2=z\sqrt{2}(J_c-J), \quad \mu_r=\mu-\mu_c, \quad c=\sqrt{\sqrt{2}J_c}.
\end{equation}

From Eq.~\eqref{Eq:ENT2} we can see that inside the Mott-insulator phase $\Delta S_E$ will be independent of the chemical potential,
\begin{equation}
\Delta S_E/A = \frac{1}{6} \sqrt{z} \sqrt{1 - \frac{J}{J_c}}.
\end{equation}
Inside the superfluid phase in the relativistic regime ($\mu=\mu_c$) we obtain,
\begin{equation}
\Delta S_E/A = \frac{1}{12} \sqrt{2z} \sqrt{\frac{J}{J_c} - 1}.
\end{equation}

\end{document}